\documentclass[twocolumn,prl,aps,showpacs,floatfix]{revtex4}

\usepackage{dcolumn}
\usepackage{amsmath}

\newcommand{\wtR}{\widetilde{R}}
\newcommand{\tdr}{\tilde{r}}
\newcommand{\tdd}{\tilde{d}}

\newcommand{\ice}[1]{\relax}

\newcommand{\re}[1]{(\ref{#1})}

\newcommand{\unl}{\underline}

\newcommand{\as}{a_s}

\newcommand{\als}{\alpha_s}
\newcommand{\beq}{\begin{equation}}

\newcommand{\ba}{\begin{array}}

\newcommand{\ea}{\end{array}}

\newcommand{\eeq}{\end{equation}}

\newcommand{\bea}{\begin{eqnarray}}

\newcommand{\eea}{\end{eqnarray}}

\newcommand{\G}{\Gamma}

\newcommand{\EQN}{\label}

\newcommand{\cF}{{\cal F}}

\newcommand{\MeV}{\mbox{\rm MeV}}
\newcommand{\GeV}{\mbox{\rm GeV}}

\begin{document}

\title{

{
 \vspace*{-14mm}

\centerline{\normalsize\hfill  SFB/CPP-05-33}
\centerline{\normalsize\hfill  TTP05-11    }
\centerline{\normalsize\hfill hep-ph/0511063}
\baselineskip 11pt
{}
}
\vspace{3mm}

Scalar Correlator at ${\cal O}(\alpha_s^4)$, Higgs Decay into $b$-quarks and
Bounds on the Light Quark Masses  
\vskip.3cm
     }
\author{P.~A.~Baikov}
\affiliation{Institute of Nuclear Physics, Moscow State University,
Moscow~119899, Russia
        }
\author{K.~G.~Chetyrkin}\thanks{{\small Permanent address:
Institute for Nuclear Research, Russian Academy of Sciences,
 Moscow 117312, Russia}}

\author{J.~H.~K\"uhn}
\affiliation{Institut f\"ur Theoretische Teilchenphysik,
  Universit\"at Karlsruhe, D-76128 Karlsruhe, Germany}

\begin{abstract}
\vspace*{.7cm}
\noindent
We  compute, for the first time,  the absorptive part of the
massless correlator of two quark scalar currents in five loops.  As
physical applications we consider the ${\cal O}(\alpha_s^4)$
corrections to the decay rate of the Standard Model Higgs boson into
quarks, as well as the  constraints 
on the strange quark mass following  from  QCD sum rules.
\end{abstract}


\maketitle

\section{Introduction}   

Within the Standard Model (SM) the scalar Higgs boson is responsible
for the mechanism of the mass generation.  Particularly interesting for
the observation of the Higgs boson with an intermediate mass
$M_H<2M_W$ is the dominant decay channel into a pair of bottom quarks, 
$H\rightarrow b\bar{b}$.  The decay of the Higgs boson into a
quark--antiquark pair $(\bar{f}f)$ proceeds through its coupling
to the corresponding quark scalar current and reads %
(for a review  see e.g. \cite{Djouadi:2005gi}.)
\ice{
\beq
{\cal L}_Y = \sum_f m_f J^S_f, \ \ J^S_f = \bar{\Psi}_f\Psi_f
{}. 
\eeq
}
\beq 
\G(H \to \bar{f}f )
=\frac{G_F\,M_H}{4\sqrt{2}\pi}
m_f^2  \wtR (s = M_H^2)
\label{decay_rate_from_R}
{},
\eeq
with   
$
  \wtR(s)
 = 
\mbox{\rm Im} \,  \widetilde{\Pi}(-s-i\epsilon)/{(2\pi\,  s)} 
$
standing for the absorptive part of the scalar two-point 
correlator:
\beq
\EQN{Pi}
\widetilde{\Pi} (Q^2)
= (4\pi)^2 i\int dx e^{iqx}\langle 0|\;T[\;J^{\rm S}_f(x)
J^{\rm S}_{f}(0)\,]\;|0\rangle
{}.
\eeq
Here $Q^2 = - q^2$ and  $J^{\rm S}_f=\bar{\Psi}_f\Psi_f$ is the scalar
current for  quarks with flavour $f$ and mass $m_f$,  coupled
to the scalar Higgs boson.

The currently known fixed order perturbative  predictions for $\widetilde{R}$
can be shortly summarized as follows \cite{Gorishnii:1990zu,gssql4} 
(we have put the number of  effective flavours $n_f=5$)
\beq
 \widetilde{R}= 1 + 5.66667\,  a_s+ 29.1467 \, a_s^2  +  41.7576 \,a_s^3  
\label{RS_as3}
{},
\eeq
with   $\as = \alpha_s(M_H)/\pi$.
Note that eq.~\re{RS_as3} is given in the high energy limit with all 
power-suppressed terms of order $m_f^2/M^2_H$ and higher neglected. In fact,
the full dependence on the quark mass $m_f$ is known up to and
including the ${\cal O}(\alpha_s^2)$ contribution \cite{Chetyrkin:1997mb}. 
We will not discuss the power suppressed terms in
the present publication.

For  $M_H \ge 70$ GeV 
already the term of order $m_b^2/M_H^2\, \alpha_s^2 $ is
numerically by an order of magnitude less than the massless 
${\cal O}(\alpha_s^3)$ term displayed in eq.~\re{RS_as3} \cite{gssql4}.
Due to the large mass $M_H$ the couplant $\as $ is less than $0.04$ which results 
in a good apparent convergence of the perturbation  series in eq.~\re{RS_as3}.

On the other hand, for  energy scales, say, 
of order of a few GeV's, relevant for  QCD sum rules  the higher order corrections to
the correlator \re{Pi} are numerically  quite important (see,
e.g. \cite{Chetyrkin:1997xa,Maltman:2001gc,Jamin:2001zr,Narison:2002hk}). 
We will discuss this issue in some detail later in the text for an
example of bounds for the light quark masses.

The motivation of the present publication is fourfold. First,
presenting the {\em first complete } ${\cal O}(\alpha_s^4)$ result for
a massless QCD correlator shows that  the  new theoretical methods used in
\cite{ChBK:vv:as4nf2,ChBK:tau:as4nf2,ChBK:vv:mq2as4nf2} 
indeed do deliver genuine QCD results in five-loop
approximation. Second, the results are important for the QCD sum rules
based on a \mbox{(pseudo)-scalar} correlator \re{Pi}; they also
provide an accurate prediction of the Higgs decay rate into hadrons.
Third, the case of the scalar correlator should be considered   a
necessary preparation step before computing the ${\cal O}(\alpha_s^4)$
contribution to the vector correlator. The importance of the latter
calculation for the precise determination of the value of $\alpha_s$
from the $\tau$-lepton and $Z-$boson decay rates is well-known.  
Last, by comparing the exact $\als^4$ result with the estimates based
on various optimization procedures one obtains important insights into
the quality of different approaches, confirming some and refuting
others.

\section{Calculation and Results}
To compute the absorptive part  of $\widetilde{\Pi}$ we proceed in full
analogy with our previous calculations described in \cite{Baikov:2004tk}. 
First, using the criterion of irreducibility of Feynman integrals
\cite{Baikov:criterion:00,Baikov:2005nv}, 
the set of irreducible integrals involved
in the problem was constructed.
Second, the coefficients multiplying
these integrals were calculated as series in the $1/D\rightarrow0$
expansion with the help of an auxiliary integral representation 
\cite{Baikov:tadpoles:96}. Third, the exact answer, i.e.  a rational function of $D$,
was reconstructed from this expansion.  

The major part of the calculations was performed  on 
the Silicon Graphics Altix 3700 computer (32 Itanium-2 1.3 GHz processors) 
using the parallel version of FORM 
\cite{Vermaseren:2000nd,Fliegner:1999jq,Fliegner:2000uy} 
and  took about 18 months  in total.
The  diagrams were generated with the help of QGRAF \cite{Nogueira:1991ex}.

It is convenient to introduce  the Adler function as
\begin{equation}
\widetilde{D}(Q^2) =  \frac{Q^2}{6}
\frac{\mathrm{d}}{\mathrm{d} Q^2} \frac{\widetilde{\Pi}(Q^2)}{Q^2}
=
\int_0^\infty \frac{ Q^2\ \widetilde{R}(s) d s }{(s+Q^2)^2}
{}\ ,
\label{Adler_scalar}
\end{equation}
\beq
\nonumber
 \widetilde{D}(Q^2) = 1+ \sum_{i=1}^{\infty} \  \widetilde{d}_i a_s^i(Q^2),  \ \
\tilde{R}(s) = 1 + \sum_{i=1}^{\infty} \  \tilde{r}_i a_s^i(s)
{}\ ,
\eeq
where we have set the normalization scale $\mu^2=Q^2$ or to $\mu^2= s$
for the Euclidian and Minkowskian functions respectively.  
The specific form  chosen in eq.~(\ref{Adler_scalar})
is particularly convenient for perturbative  consideration.
The results
for generic values of $\mu$ can be easily recovered with the standard
RG techniques. Since the functions  $\tilde{D}$
is  known to order $\alpha_s^3$ from
\cite{gssql4} we   only display the
${\cal O}(\alpha_s^4)$ results.
\begin{widetext}
\begin{eqnarray}
\lefteqn{\tilde{d}_4 =  } 
\nonumber\\
&{+}& \, n_f^3
\left[
-\frac{520771}{559872} 
+\frac{65}{432}  \,\zeta_{3}
+\frac{1}{144}  \,\zeta_{4}
+\frac{5}{18}  \,\zeta_{5}
\right]
\nonumber\\
&{+}& \, n_f^2
\left[
\frac{220313525}{2239488} 
-\frac{11875}{432}  \,\zeta_{3}
+\frac{5}{6}  \,\zeta_3^2
+\frac{25}{96}  \,\zeta_{4}
-\frac{5015}{432}  \,\zeta_{5}
\right]
\nonumber\\
&{+}& \, n_f 
\left[
-\frac{1045811915}{373248} 
+\frac{5747185}{5184}  \,\zeta_{3}
-\frac{955}{16}  \,\zeta_3^2
-\frac{9131}{576}  \,\zeta_{4}
+\frac{41215}{432}  \,\zeta_{5}
+\frac{2875}{288}  \,\zeta_{6}
+\frac{665}{72}  \,\zeta_{7}
\right]
\nonumber\\
&{+}&
\left[
\frac{10811054729}{497664} 
-\frac{3887351}{324}  \,\zeta_{3}
+\frac{458425}{432}  \,\zeta_3^2
+\frac{265}{18}  \,\zeta_{4}
+\frac{373975}{432}  \,\zeta_{5}
-\frac{1375}{32}  \,\zeta_{6}
-\frac{178045}{768}  \,\zeta_{7}
\right]
{}.
\label{NPsip*QQ}
\end{eqnarray}
\end{widetext}
The $n_f^3$ and $n_f^2$ terms  are  in agreement  with 
\cite{Broadhurst:2000yc}  and \cite{ChBK:vv:as4nf2}
respectively.
Once the constants $\tilde{d}_1$ to $\tilde{d}_4$ are known it is  a
matter of  straightforward analytic continuation to find $\widetilde{R}$,
given below for brevity in the numerical form  only:
\bea
\widetilde{R} &=& 
1
+
5.6667  a_s
{+} 
\left[
35.94
-1.359  \, n_f
\right]
a_s^2
\nonumber\\
&{+}& a_s^3
\left[
164.14
-25.77  \, n_f
+0.259  \, n_f^2
\right]
\label{RSnum}\\
&{+}& \,a_s^4
\left[
39.34
-220.9  \, n_f
+9.685  \, n_f^2
-0.0205  \, n_f^3
\right]
{}.
\nonumber
\eea

In order to better understand  the structure of the $\alpha_s^4$ term in
\re{RSnum} it is  instructive to separate  the genuine five-loop  contributions from 
the Adler function $\widetilde{D}$ from  essentially
``kinematical'',  so-called $\pi^2$-terms  originating  from the analytic continuation.
We have in mind that for a given order in
$\alpha_s$ these extra contributions are completely predictable from
the standard evolution equations  applied to the ``more leading'' terms in $\widetilde{D}$
proportional to  some smaller   powers of   $\alpha_s$.  
The corresponding expression for $\widetilde{R}$ reads
\ice{
The corresponding expressions
for the coefficients  $\tilde{r}_2$ --- $\tilde{r}_4$ read 
}
\ice{
\begin{eqnarray}
\tilde{r}_2 &=& 51.57 - \unl{15.63}, 
\\
\tilde{r}_3 &=&
{648.7} - \unl{484.6} 
- \, n_f ({63.74}  - \unl{37.97})
+ n_f^2 ({0.929}  \,
-\unl{0.67}) 
\\
\tilde{r}_4 &=&  9471.0 - \unl{9431.0}
- n_f (1454.0   - \unl{1233.0})
+  \, n_f^2 (54.78 - \unl{45.1})
- \, n_f^3 (0.454 -\unl{0.433})
\end{eqnarray}
}
\begin{widetext}
\begin{eqnarray}
\widetilde{R} &=&  
{1}
{+} 
{5.667} \as
{+} a_s^2
\left[
51.57 - \unl{15.63} 
- \, n_f (
 {1.907} 
- \unl{0.548} 
)
\right]
{+} a_s^3
\left[
{648.7} - \unl{484.6} 
- \, n_f ({63.74}  - \unl{37.97})
+ n_f^2 ({0.929}  \,
-\unl{0.67})
\right]
\nonumber\\
&{+}& \,a_s^4
\left[
9470.8 - \unl{9431.4}
- n_f (1454.3   - \unl{1233.4})
+  \, n_f^2 (54.78 - \unl{45.10})
- \, n_f^3 (0.454 -\unl{0.433})
\right]
{},
\label{RSnumx}
\end{eqnarray}
\end{widetext}
where  we have underlined  the contributions coming from analytic continuation.

The inclusion of the $\pi^2$-terms from higher orders thus
does not necessarily improve the quality of the perturbative prediction for the
scalar correlator. It remains to be seen whether  a similar pattern arises in the case 
of the contour improved perturbation theory
\cite{Pivovarov:1991rh,LeDiberder:1992te} 
applied  to the $\tau$-lepton decay rate.

At last, specifying $n_f=5$ we get the corresponding  generalization of
eq.~\re{RS_as3}
\bea
\widetilde{R} &=& 1 + 5.6668\,  a_s+ 29.147 \, a_s^2  +  41.758 \,a_s^3 -   825.7\,a_s^4
\nonumber
\\
&=& 
1 + 0.2075  + 0.0391  + 0.0020   - 0.00148 
\label{RS_as4_nl5}
{}.
\eea
In the last equation we have substituted $\as= 0.0366$ which
corresponds the Higgs mass value $M_H = 120\  \GeV$.  The comparable
sizes of the ${\cal O}(\as^3)$ and the ${\cal O}(\as^4)$ terms (the
fourth and the fifth terms in eq.~ \re{RS_as4_nl5}) may be 
interpreted as a consequence of the accidentally small coefficient for
the $\as^3$ term.

\section{Comparisons to Previous Estimations}
Following \cite{Kataev:1995vh},  predictions for higher order
coefficients of QCD correlators are usually made first for an
Euclidian quantity and  the result for the corresponding
Minkowskian one is then derived as a direct consequence.
On the other hand, a  quite remarkable feature of our result for $\wtR$  are  almost complete
cancellations which take  place for every power of $n_f$ 
between the genuine five-loop contributions 
and the ``trivial'' ones from analytic continuation as illustrated
by the corresponding decomposition  displayed in eq.~\re{RSnumx}. This fact alone 
might  indicate   problems for the traditional way of obtaining  predictions for
$\wtR$. With the exact  result in hand one  can easily check this suspicion.

Indeed,  in Table \re{tab:FACPMS}  our results 
are compared with predictions obtained in works
\cite{CheKniSir97,Chishtie:1998rz,Broadhurst:2002bi}.
Note  that the Principle of Minimal Sensitivity (PMS) 
\cite{Stevenson:1981vj} and that of Fastest Apparent Convergence (FAC) 
\cite{Grunberg:1982fw}  used in
\cite{CheKniSir97} produce  identical result at order $\alpha_s^4$. 
Also note that the Asymptotic Pad\'e-Approximant Method (APAM)
was utilized in \cite{Chishtie:1998rz} to produce the prediction directly
(and only) for the absorptive part $\wtR$. In contrast,  the method of Naive 
NonAbelianization (NNA)  \cite{Broadhurst:1995se} has been applied  in  
\cite{Broadhurst:2002bi} to  the Adler function only.
The two predictions of FAC/PMS
 for $\tdr_4$ correspond to either the  consequence of the
prediction for $\tdd_4$ (the fifth line) or to the direct application
of FAC/PMS to estimate $\tdr_4$ (the sixth line).  
As a  consequence of the large cancellations in $\tdr_4$ 
the second prediction looks much better than the first, despite  the fact that
the estimation of the Euclidian coefficient $\tdd_4$ is quite close
(within 10\%) to the exact result. In fact, at order  $\alpha_s^3$   the
cancellations in questions are much less pronounced with the  result that
the corresponding prediction for $\tdr_3$,  obtained from $\tdd_3$, is
significantly more accurate than the direct estimation of $\tdr_3$
\cite{CheKniSir97}. NNA predictions have  correct signs and sensible
magnitudes as observed earlier \cite{Broadhurst:2002bi}.
Finally, the APAM estimation of
$\tdr_4$  fails to reproduce  even the sign of the exact result.

Predictons of the prescription proposed by Brodsky, Mackenzie and Lepage (BLM)
\cite{Brodsky:1982gc} for the $n_f$ dependent terms of order $\alpha_s^4$ have been communicated 
to the authors \footnote{M.~Binger and  S.~Brodsky, private commication.}:
$\as^4( -260 \, n_f + 13\, n_f^2  - 0.046\,  n_f^3)$
and are also in reasonable  agreement with the exact result of eq.~(\ref{RSnum}).

\begin{table}[h]
\caption{\label{tab:FACPMS}Comparison
        of the results obtained in this
        paper with earlier estimates based on
        PMS, FAC  and APAM.
         }
\begin{ruledtabular}
\begin{tabular}{cccc}
$n_f$                      & 3       & 4       &     5  \\
\hline\\
$\tdd_4$ (exact) & 5588.7  & 4501.1  & 3512.2 \\
$\tdd_4$ (\cite{CheKniSir97}, PMS, FAC)   & 5180.3  &   4093.0  & 3100.5  \\
$\tdd_4$ (\cite{Broadhurst:2002bi}, NNA)   & 1592.8  &   1521.4  & 1484.1  \\
\\
\hline\\
$\tdr_4$ (exact)                          & -536.8  &  -690.7   & -825.7  \\
$\tdr_4$ (\cite{CheKniSir97}, PMS, FAC)   & -945.2  &  -1098.8  & -1237.4 \\
$\tdr_4$ (\cite{CheKniSir97}, PMS, FAC)   & -527.8  &  -748.6  & -949.4 \\
$\tdr_4$ (\cite{Chishtie:1998rz}, APAM)             &  252    &  147      & 64.2  \\
\end{tabular}
\end{ruledtabular}
\end{table}


\section{Quark Mass Bounds}

As an application of our result for the scalar correlator we consider
the well-known bounds for the light quark masses
\cite{Becchi:1980vz,Lellouch:1997hp}. The constraints follow from the
known values of the $\pi$ or $K$ pole contributions and the positivity
of the spectral function $\wtR$ and depend on the scale $Q$ used in
evaluation of the polarization operator.
We are going to discuss two types of bounds suggested in \cite{Lellouch:1997hp}, the ``quadratic'' bound 
\begin{widetext}
\beq
\left[m_{s}(Q)+m_{u}(Q)\right]^2 \ge
\frac{16\pi^2}{N_c}
\frac{2f_{K}^2
M_{K}^4}{Q^4}\frac{9}{\left(1+\frac{M_{K}^2}{Q^2}\right)^5}
\frac{2\cF_{0}(Q^2) -\frac{4}{3}
\left(1+\frac{M_{K}^2}{Q^2}\right)\cF_{1}(Q^2)
+\frac{1}{3}\left(1+\frac{M_{K}^2}{Q^2}\right)^2\cF_{2}(Q^2)}
{3\cF_{0}(Q^2)\cF_{2}(Q^2)
-2\left(\cF_{1}(Q^2)\right)^2}
\label{quadratic}
\eeq
\end{widetext}
and the ``linear'' bound 
\beq
\left[m_{s}(Q)+m_{u}(Q)\right]^2 \ge \frac{16\pi^2}{N_c}
\frac{2f_{K}^2 M_{K}^4}{Q^4}
\frac{\left(1+\frac{M_{K}^2}{Q^2}\right)^{-3}}{\cF_{0}}
{}.
\label{linear}
\eeq
Here the functions  $\cF_0,\cF_1$ and $\cF_2$ are defined as (normalized to one at the leading order, 
that is $c_0= 1/6, c_1 =-1/6, \ c_2 = 1/12$) 
\bea
\cF_{n}(Q^2)= c_n\  (Q^2)^{1+n}\, \left(\frac{\mathrm{d}}{\mathrm{d} Q^2}\right)^{2+n} 
\widetilde{\Pi}(Q^2)
\label{Fn}
{}.
\eea
The results of both eqs.~(\ref{quadratic},\ref{linear}) for $m_s+m_u$
depend on the choice of $Q$  and  can be transformed to  bounds for
$(m_s+m_u)(\mu)$ for any (not too small) $\mu$ using the standard mass
evolution equation. Following \cite{Lellouch:1997hp} we will use  $\mu = 2\  \GeV$
as the reference point. 

Let us start from \re{quadratic}. 
Using as reference value $\alpha_s(M_{\tau}) = 0.334$ 
and the standard  evolution  equations for the mass and coupling constant   we get
(the indices 4 and 3  stand   for the order in $\alpha_s$)
\beq
[(m_s+m_u)^2]_4 \left(
\ba{c}{\scriptstyle \mu=2} 
\\ 
{\scriptstyle Q = 2}
 \ea 
\, \GeV \right)  
> (103  \   \MeV)^2
\label{bound_4}
{},
\eeq
which should be compared to the three-loop bound \cite{Lellouch:1997hp}
\beq
[(m_s+m_u)^2]_3 
\left(
\ba{c} {
\scriptstyle \mu=2} 
\\ 
{\scriptstyle Q = 2}
 \ea 
\, \GeV \right)  
> (111 \   \MeV)^2
\label{bound_3}
{}.
\eeq

The bound \re{bound_3}, if valid, is already in conflict with significantly  lower
value 75(8) MeV  derived  by one of lattice collaborations \cite{Aubin:2004ck}.
(However, for a significantly larger  lattice 
result around   a hundred of MeV, see \cite{Gockeler:2004rp,DellaMorte:2005kg}).

The problem  was ``solved'' in 
\cite{Maltman:2001gc}  by observing that the quadratic combination 
of the $\cF_{i}$ functions appearing in the 
denominator of \re{quadratic} displays a poor pattern of convergence:
\beq
3 {\cal F}_0 {\cal F}_2  - 2 ({\cal F}_1)^2 = 
1  + 0.83 + 0.61 + 0.51 + \ldots 
\eeq
and then suggesting to increase $Q$ in \re{bound_3} till, say, $2.5 \  \GeV$,  which
lowers the rhs of  \re{bound_3} to $86 \ \MeV$ and the rhs of  \re{bound_4} 
to $81 \ \MeV$. 

From our point of view it is more relevant to consider the convergence
pattern of the whole ratio appearing in the rhs of eq.~\re{quadratic}. 
Indeed, after expanding \re{quadratic} in $\as$ and
disregarding all terms in $\as$ higher than $\as^4$ we get
\bea
&{}&
[(m_s+m_u)^2]_4 \left(
\ba{c}{\scriptstyle \mu=2}
\\ 
{\scriptstyle Q = 2}
 \ea 
\, \GeV \right)  
> (179  \   \MeV)^2 
\times
\nonumber
\\
&{}&
\left\{
1 - 6.44 \,\as -12.83\, \as^2 + 
(482.95 - \unl{525.2})\, \as^3 
\nonumber
\right.
\\
&{}&
\left.
\phantom{X} + (-2561.8 + \unl{6948.34} - \fbox{4439.9})\, \as^4  
\right\}
\label{bound_4_expanded}
{}.
\eea

In order to demonstrate the importance  of higher order corrections 
we have underlined in \re{bound_4_expanded} all terms originating from  
the  contributions of order $\as^3$ 
(eventually multiplied  by a subleading ${\cal O}(\as)$ term) 
to the functions  $\cF_0,\cF_1$ and $\cF_2$ and, correspondingly,
boxed the one
originating  exclusively from  ${\cal O}(\as^4)$ terms in  the same functions. 
After substituting the value for $\alpha_s(2 \ \GeV) = 0.312$ 
into eq.~\re{bound_4_expanded}
we arrive at 
\beq 
[(m_s+m_u)]_4 (\mu =2\, \GeV)  > 77 \ \MeV  
\ice{\ (\mbox{\re{quadratic} expanded to}\  \as^4)}
{} 
\label{4l}
\eeq
to be compared to 
\beq 
[(m_s+m_u)]_3 (\mu =2\, \GeV)  > 79 \ \MeV
\ice{(\mbox{\re{quadratic} expanded to}\  \as^3)}
\label{3l}
{}. 
\eeq
If, on the other hand, we would try to use eq.~\re{bound_4_expanded}
with only the boxed term nullified we would  get an astonishingly different value
\beq 
[(m_s+m_u)]_3(\mu =2\, \GeV)  > 141 \ \MeV 
\label{3land4l}
\ice{(\mbox{ \re{quadratic} expanded to}\  \as^4)}
{}. 
\eeq

In similar way one  can examine the second bound  \re{linear}. 
For the lowest  choice
$Q =1.4\  \GeV$ used in \cite{Lellouch:1997hp} we immediately
get 
\beq
[(m_s+m_u)](\mu =2\, \GeV) > 76 \  \MeV \ \mbox{and} \ 78 \ \MeV
\nonumber
{},
\eeq
with the   numbers corresponding to the use of five- and four-loop 
expressions  for the function $\cF_{0}$ correspondingly. 

On the other hand, the expanded version of   eq.~\re{linear}
reads:
\bea
&{}&
[(m_s+m_u)^2]_4 \left(
\ba{c}{\scriptstyle \mu=2}
\\ 
{\scriptstyle Q = 1.4}
 \ea 
\, \GeV \right) 
>
 (106 \   \MeV)^2 
\times
\nonumber
\\
&{}&
\left\{
1 - 3.67 \,\as - 0.73\, \as^2 + 
(54.7 - \unl{77.4})\, \as^3 
\nonumber
\right.
\\
&{}&
\left.
\phantom{X} + (-190.1 + \unl{567.4} - \fbox{511.8})\, \as^4  
\right\}
\label{bound_linear_expanded}
{}.
\eea
It  leads to the following bounds
\beq
[(m_s+m_u)]_4 (\mu =2\, \GeV)  > 72 \ \MeV  
\ice{\ (\mbox{\re{quadratic} expanded to}\  \as^4)}
{}, 
\label{4l_linear}
\eeq
\beq 
[(m_s+m_u)]_3 (\mu =2\, \GeV)  > 74 \ \MeV
\ice{(\mbox{\re{quadratic} expanded to}\  \as^3)}
\label{3l_linear}
{}. 
\eeq
Again, if we were using use eq.~\re{bound_linear_expanded}
with only the boxed term nullified we would arrive at significantly 
different value
\[
[(m_s+m_u)]_4 (\mu =2\, \GeV)  > 81 \ \MeV  
{}.
\]

\section{Conclusion}

We have computed the correction of order $\alpha_s^4$ to the
correlator of quark scalar currents in the massless limit. Our result
demonstrates a remarkable interplay between the genuine five-loop terms
and the effects due to the analytical continuation. The
newly computed contribution stabilizes the (quadratic) quark mass
bound of \cite{Lellouch:1997hp}  and pushes it significantly down,  thus, 
avoiding any potential conflict with  lattice results.

The authors are grateful to the members of the 
Institut f\"ur Theoretische Teilchenphysik of the Karlsruhe
University (especially to Michael Faisst and Michail Tentyukov) for all
their effort to provide the flawless running of the Silicon Graphics
cluster, which was critical for completing these calculations. 
We thank Rainer Sommer for a usefull discussion. 

This work was supported by 
the Deutsche Forschungsgemeinschaft in the 
Sonderforschungsbereich/Transregio
SFB/TR-9 ``Computational Particle Physics'',  by INTAS (grant
03-51-4007) and by RFBR (grants 03-02-17177  and 05-02-17645).

\end{document}